\documentclass[aps,prd,reprint,preprintnumbers,nofootinbib]{revtex4-1}

\usepackage{amsmath}
\usepackage{amsfonts} 
\usepackage{amssymb}
\usepackage{graphicx}
\usepackage{hyperref}
\usepackage{tensor}
\usepackage{siunitx}
\usepackage{float}
\usepackage{bm}
\usepackage{mathtools}
\DeclarePairedDelimiter{\floor}{\lfloor}{\rfloor}
\newcommand{\be}{\begin{equation}}
\newcommand{\ee}{\end{equation}}

\begin{document}

\title{\Large On Calabi-Yau manifolds at strong topological string coupling}

\author{\large Jarod Hattab}
\author{\large Eran Palti}

\affiliation{\vspace{0.4cm}
Department of Physics, Ben-Gurion University of the Negev, Beer-Sheva 84105, Israel}

\begin{abstract}
It was recently shown that integrating out M2 states on Calabi-Yau manifolds captures non-perturbative topological string physics in the free energy. In this note, we show that the resulting expression manifests a certain duality symmetry: the free energy at strong string coupling is equal to the Calabi-Yau period at weak string coupling. The duality yields the appropriate prescription for completing the integrating out in the ultraviolet.
\vspace{1cm}
\end{abstract}

\maketitle

\section{Introduction}
\label{sec:int}

Recently, in \cite{Hattab:2024ewk}, it was shown that integrating out M2 branes on Calabi-Yau manifolds yields non-perturbative contributions to the free energy for closed topological strings. It remains to be seen if these are all the non-perturbative contributions, but it was conjectured in \cite{Hattab:2024ewk} that it captures all the ``Wilsonian'' contributions.\footnote{There are additional contributions associated to degenerate target space worldsheet geometries, which relate to the so-called holomorphic anomaly. In the supergravity interpretation, these are (non-local) non-Wilsonian contributions \cite{Dedushenko:2014nya}.} In this note, we show that the non-perturbative result manifests a certain duality symmetry. The duality exchanges the Calabi-Yau prepotential (topological string free energy) at strong coupling with the Calabi-Yau period at weak coupling. 

S-duality for topological strings was studied already in \cite{Neitzke:2004pf,Nekrasov:2004js}, where it was proposed to relate the A and B models on the same Calabi-Yau. Further, it was argued to support a picture of the A model as a melting crystal \cite{Iqbal:2003ds,Nekrasov:2004js,Ooguri:2009ijd}. The duality relation we discuss bears similarities to these ideas, but it is more precisely matched by the results in \cite{Drukker:2010nc,Marino:2011eh,Grassi:2014zfa}.\footnote{See also, for example, \cite{Pasquetti:2010bps,Hatsuda:2013oxa,Couso-Santamaria:2013kmu,Couso-Santamaria:2014iia,Gu:2023mgf,Alexandrov:2023wdj,Alim:2024dyi} for work and \cite{Marino:2012zq,ANICETO20191} for reviews, on studies using Borel resummation and resurgence relating non-perturbative contributions to periods.} In particular, in \cite{Grassi:2014zfa} the precise form of our duality was shown for a number of non-compact toric examples. It was suggested that it should hold for general toric non-compact Calabi-Yau manifolds. It was subsequently conjectured in \cite{Hattab:2024thi} to hold generally, including for compact Calabi-Yau manifolds. This was proposed from an integrating out motivation, interpreting the period as an integrating out calculation. In this note we show that the results in \cite{Hattab:2024ewk} confirm the conjecture of \cite{Hattab:2024thi}.

For simplicity, here and henceforth we consider Calabi-Yau manifolds with a single (complexified) Kahler modulus $t$.
The duality occurs after a change of variable which exchanges $t$ for a parameter $\mu$ as
\be
\label{Ttomu}
t = \frac{2\pi\mu}{\hbar} \;\;,\;\;\;\; \hbar \equiv \frac{\left(2\pi\right)^2}{\lambda} \;,
\ee
with $\lambda$ being the topological string coupling. Such a relation appears in \cite{Marino:2011eh} where topological strings on the blown-up conifold were shown to be equivalent to a free Fermi gas. In that setting, $\mu$ was a chemical potential associated to the Fermion number $N$, which in turn, was equal to the $N$ of the dual $SU(N)$ Chern-Simons theory. Since the Chern-Simons theory localises to a matrix model, for those settings, $\mu$ is a chemical potential on an ensemble of matrix theories. 


The duality is then as follows. Consider a weak string coupling expansion of the topological string free energy, the usual genus expansion, 
\be 
\label{Fexp}
F\left(t\right) = \sum_{g=0}^{\infty} F_g\left(t\right) \lambda^{2g-2}\;.
\ee 
Then the Calabi-Yau period is defined from the genus-zero prepotential as
\be 
\label{perioddef}
\Pi_0\left(t\right) = 2 F_0\left(t\right) - t \frac{\partial F_0\left(t\right)}{\partial t}\;.
\ee 
We then take $\Pi_0\left(t\right)$ and send $t \rightarrow \mu$, so define the period as a function of $\mu$ through 
\be
\label{perinmu}
\Pi_0\left(\mu\right) = \Pi_0\left(t\right) \big|_{t = \mu} \;.
\ee
Next we consider the free energy after the change of variable (\ref{Ttomu}). So we define
\be 
\label{muTfree}
F\left( \mu \right) =  F\left(t\right) \big|_{t = \frac{2\pi\mu}{\hbar}} \;.
\ee 
We then expand this about strong coupling  
\be 
\label{Fmuexpan}
F\left(\mu\right) = \sum_{n=0}^{\infty} J_n\left(\mu\right) \left(2\pi \hbar\right)^{2n-1} \;.
\ee 
Then the duality relation is 
\be
\label{cyper}
J_0\left( \mu \right) = -\Pi_0 \left( \mu \right) \;.
\ee
In this note, we show that (\ref{cyper}) follows from the expression for the free energy arising from integrating out M2 states derived in \cite{Hattab:2024ewk}. We show that this is true for any Calabi-Yau manifold, as conjectured in \cite{Hattab:2024thi}.\footnote{Note that in \cite{Hattab:2024thi} the relation is without the minus sign. This is because $F_0$ can be defined as an expansion in the string coupling $g_s$ or as an expansion in the topological string coupling $\lambda$. The two are related by $g_s = i\lambda$, which flips the sign of $F_0$.}

In section \ref{sec:per}, we discuss how the relation to the period (\ref{cyper}) determines the prescription for the ultraviolet completion of the integrating out calculation. This follows from the existence of an integral representation for the period. The prescription shows that the ultraviolet contribution to integrating out yields the classical (weak-coupling) tree-level prepotential terms. This relation is what was termed Dual Emergence in \cite{Hattab:2024chf}, as a refinement of the Emergence Proposal \cite{Palti:2019pca} (based on ideas in \cite{Harlow:2015lma,Heidenreich:2017sim,Grimm:2018ohb,Heidenreich:2018kpg,Palti:2019pca}), and matches the results of \cite{Hattab:2023moj,Hattab:2024thi,Hattab:2024ewk,Hattab:2024chf}.\footnote{See also \cite{Blumenhagen:2023tev,Blumenhagen:2023xmk,Blumenhagen:2024ydy,Blumenhagen:2024lmo} for a closely related approach to emergence.} 

We also show, in appendix \ref{sec:part}, that the expression in \cite{Hattab:2024ewk} for the free energy derived from integrating out through a Schwinger computation can be reproduced from a particle worldline computation, as considered in \cite{Dedushenko:2014nya}.

\section{Strong coupling expansion}
\label{sec:stcpex}

The general result for the topological string free energy from integrating out M2 states was given in \cite{Hattab:2024ewk} as
\be 
\label{full}
   \left. F\left(t\right) \right|_{\mathrm{Exp}} = \sum_{{\bf \beta},r\geq 0}\alpha_r^{{\bf \beta}}\oint_C\frac{\text{d}u}{u}\frac{1}{1-e^{-2\pi i u}}\frac{e^{-\left(\beta t-2\pi i n_{\beta b} \right)u}}{\big(2\sin(u\lambda/2)\big)^{2-2r}} \;.
\ee 
Here we restricted the general expression in \cite{Hattab:2024ewk} to a single Kahler modulus $t$. The contour integration path $C$ goes around the positive real axis (for $\lambda \in \mathbb{R}^+$) but excludes the zero point. The sum is over the internal wrapping number of the M2 states $\beta$, the genus $r$, and the $\alpha^\beta_r$ are the Gopakumar-Vafa invariants counting the degeneracy of the BPS states. Finally, we have $n_{\beta b} = \floor{\frac{\beta b}{2\pi}} \in \mathbb{Z}$, with $b=\mathrm{Im\;} t$. 

The restriction $|_{\mathrm{Exp}}$ in (\ref{full}) means one keeps only the parts in $F(t)$ which are exponential in $t$. There are also polynomial parts which are dropped. Usually, this is not specified explicitly, but for our purposes it is important to keep track of it. 

The result (\ref{full}) is a modification of the seminal calculation of Gopakumar and Vafa \cite{Gopakumar:1998ii,Gopakumar:1998jq}, which was also studied in more detail in \cite{Dedushenko:2014nya}. In particular, \cite{Dedushenko:2014nya} presented a particle worldline approach to the calculation. This suggests that we should be able to rederive (\ref{full}) from that approach too. We show how to do this in appendix \ref{sec:part}.

We can now perform the change of variable (\ref{Ttomu}), and rescale $u \rightarrow \frac{u \hbar}{2\pi}$, to give
\be 
\label{rcresmu}
\left.F(\mu) \right|_{\mathrm{Exp}} =\sum_{{\bf \beta},r\geq 0}\alpha_r^{{\bf \beta}}\oint_C\frac{\text{d}u}{u}\frac{1}{1-e^{-i \hbar u}}\frac{e^{-\beta \mu u}}{\big(2\sin(\pi u)\big)^{2-2r}} \;,
\ee 
where we have set $n_{\beta b}=0$ for simplicity. 
Note that in terms of $\hbar$ the perturbative and non-perturbative poles are interchanged. So that now the poles from the inverse powers of $\sin \left(\pi u\right)$ yield perturbative terms in $\hbar$, while the poles at $u=\frac{2\pi}{\hbar} \mathbb{N}^*$ yield non-perturbative terms in $\hbar$. 

We can extract the leading terms in $\hbar$, as in the expansion (\ref{Fmuexpan}), yielding
\be 
\label{j0intrep}
\left.J_0\left(\mu\right)\right|_{\mathrm{Exp}} = \oint_C du \left( \sum_{{\bf \beta}}\alpha_0^{{\bf \beta}} \frac{2\pi}{iu^2}\frac{e^{-\beta\mu u}}{(2\sin(\pi u))^{2}} \right) \;.
\ee 
Note that this is an expression as part of an asymptotic expansion in $\hbar$, similarly to how the $\lambda$ expansion is defined. There are also non-perturbative contributions to this expansion, as captured in (\ref{rcresmu}).
It is simple to check that this satisfies the duality relation (\ref{cyper}). We can first calculate the genus zero prepotential from (\ref{full}) as
\be 
\left. F_0(t) \right|_{\mathrm{Exp}}=\oint_C du \left(  \sum_{{\bf \beta}}\alpha_0^{{\bf \beta}} \frac{1}{u^3} \frac{e^{-\beta t u}}{1-e^{-2\pi i u}} \right) \;.
\ee 
Then the period,  defined as in (\ref{perioddef}), and after sending $t \rightarrow \mu$ as in (\ref{perinmu}), is given by 
\be 
\label{pi0intrep}
\left. \Pi_0(\mu) \right|_{\mathrm{Exp}}= \oint_C du \left(  \sum_{{\bf \beta}}\alpha_0^{{\bf \beta}} \frac{\left(2 + \mu \beta u\right)}{u^3} \frac{e^{-\beta \mu u}}{1-e^{-2\pi i u}} \right) \;.
\ee 
It is simple to check that (\ref{j0intrep}) is minus (\ref{pi0intrep}).

Note that because of the restrictions $|_{\mathrm{Exp}}$, we have shown (\ref{cyper}) only for the exponential parts. Equality of the polynomial parts is simple to show, and is done in section \ref{sec:backweak}.

\section{Resummation and ultraviolet completion at strong coupling}
\label{sec:per}

The duality relation (\ref{cyper}) yields a prescription for the ultraviolet completion of the integrating out procedure. So the ultraviolet completion of (\ref{j0intrep}). This was already described in \cite{Hattab:2023moj,Hattab:2024thi,Hattab:2024ewk,Hattab:2024chf}, and in this section we expand on it and show some additional details and results. 

The formula for the free energy (\ref{full}), and therefore the resulting (\ref{j0intrep}), follows from a Schwinger one-loop integrating out calculation. The parameter $u$ is mapped to the Schwinger proper time, such that $\left|u\right| \rightarrow 0$ corresponds to short time paths and so to the ultraviolet. The removal of the zero point in the integration path $C$ is therefore an ultraviolet cutoff on the calculation. In order to remove this cutoff, we need to extend the contour to include the $u=0$ point. We define this contour as $C_0$, so $C_0 = C + \left\{0\right\}$.

We may naively try to extend the contour to $C_0$ directly in (\ref{j0intrep}) (or even in (\ref{full})). However, this is divergent in general, and does not yield the correct result. The reason is that we first must rearrange the degrees of freedom appropriately to match the description of the theory in the ultraviolet. This rearrangement amounts to a resummation which takes the schematic form
\be 
\sum_{\beta,k} ...\; e^{-\beta k \mu}\rightarrow \sum_n ... \; e^{-n \mu} \;.
\label{resumsc}
\ee 
Here the $...$ denote unspecified coefficients (which can be calculated for a given case), $\beta$ is the wrapping number of the branes, and $k$ is the pole number in the contour integral (\ref{j0intrep}), which can be equated with the wrapping number of the M-theory circle.\footnote{The resummation (\ref{resumsc}) is related to the sum over clusters in moving to the grand canonical potential in statistical mechanics in the context of \cite{Marino:2011eh}. More generally, it was proposed in \cite{Hattab:2024thi,Hattab:2024chf} that it amounts to capturing the tower of wrapped branes in a single second-quantized `brane field' that is then integrated out in the path integral.}

While the left hand side of (\ref{resumsc}) is written as a sum over contour integrals (\ref{j0intrep}), the right hand side can be written as a single contour integral. In general, this is an ambiguous statement, since we can move the sum in and out of the integral, but here it has a specific meaning because there is a standard contour integral representation of the period. Therefore, using the duality (\ref{cyper}), we can write $J_0$ in terms of this contour integral
\be 
\label{jtoiintuv}
J_0\left(\mu\right) =  -\oint_{C_0} I\left(\tilde{u}\right) \psi\left(\mu\right)^{-\tilde{u}}  \;d \tilde{u} \;. 
\ee 
Here $\psi(\mu)$ is a function which, at large $\mu$, takes the form\footnote{It is possible that some cases would have $\psi(\mu)\sim e^{p\mu}$ where $p$ is a positive integer, $p \in \mathbb{N}^*$. If we extend the integration contour over the negative poles this is necessary \cite{Hattab:2024thi}. This does not modify the discussion.}
\be
\psi\left( \mu \right) = e^{\mu} + ... \;,
\ee 
where the sub-leading terms involve higher powers of an expansion in $e^{-\mu}$.
The function $\psi\left(\mu\right)$ is determined by the mirror map.  

The form of $I\left(\tilde{u} \right)$ is well-known, and we give some examples in section \ref{sec:simre}. It has positive integer poles, which give rise to terms of the form 
\be 
\label{psimugen}
\psi(\mu)^{-n} = e^{-n\mu} + ... \;.
\ee
This is a single sum over the poles at $n$, which gives the right hand side of (\ref{resumsc}). 

The important point is that the integral representation (\ref{jtoiintuv}) is over the contour $C_0$ in $\tilde{u}$ rather than $C$. If we think of the contour integral as an integrating out calculation, so as in (\ref{j0intrep}), then the integrating out now includes the ultraviolet through the zero pole. 

We know that at least the exponential parts in $\mu$ in the integral (\ref{jtoiintuv}) can indeed be understood as an integrating out calculation, because they are precisely equal to (\ref{j0intrep}). Note that such exponential terms, in general, also arise from the zero pole in the integral (\ref{jtoiintuv}). Therefore, even parts of the zero pole have an integrating out interpretation from (\ref{j0intrep}). We therefore claim that (\ref{jtoiintuv}) is an integrating out calculation, and further, the change of variables from $e^{\mu}$ to $\psi(\mu)$ and from $u$ to $\tilde{u}$ allows us to complete the integrating out in the ultraviolet by including the zero pole. 

In (\ref{j0intrep}) the magnitude of $u$ is directly related to the energy scale being integrated out, such that $|u|\rightarrow 0$ is the ultraviolet and $|u| \rightarrow \infty$ is the infrared. After the change of variable, $|\tilde{u}|$ still has an interpretation in terms of energy scales, but the resummation (\ref{resumsc}) implies some `scale mixing' that we should explain.

 We can keep track of the energy scale by keeping track of the exponent power in $e^{-\mu}$ in the resummation (\ref{resumsc}). We note that a given $k$ can give rise to higher, but not lower, $n$ powers. In terms of $u$ and $\tilde{u}$, this is translated to the statement that physics at energy scale $|u|=u_0$ requires physics from $ |\tilde{u}| \leq u_0$ (it is necessary though not sufficient). We should therefore associate an energy scale to $|\tilde{u}|$ through the lowest power of $e^{-\mu}$ that can arise from it. In this assignment, $|\tilde{u}|\rightarrow 0$ is indeed the ultraviolet. So, in other words, capturing the ultraviolet physics of the integrating out calculation requires including the $\tilde{u}=0$ pole (the pole also gives additional infrared contributions).

Finally, since the difference between (\ref{j0intrep}) and (\ref{jtoiintuv}) are the polynomial terms, we see that including the ultraviolet physics in the integrating out gives rise to the tree-level leading polynomial terms in the free energy. 


\subsection{Examples}
\label{sec:simre}

In this section we present some examples, of increasing complexity, of $I\left(\tilde{u}\right)$ in (\ref{jtoiintuv}).

The simplest example is just the case of the Calabi-Yau being $\left(T^2\right)^3$, where the Kahler moduli of the three tori are set equal. In such a setup, there are no exponential contributions at all. Then we have a period representation 
\be 
I(\tilde{u})^{\mathrm{Torus}} =  \frac{1}{2\pi i}\frac{1}{\tilde{u}^4} \;, \;\; \psi(\mu)=e^{\mu}\;.
\ee 

The next simplest example is the resolved conifold, for which only $\alpha_0^1=1$ is non-vanishing. In this case, there is no sum over $\beta$ in (\ref{resumsc}), and so no resummation is required at all. Therefore, we can readily extend the contour integral (\ref{j0intrep}) to include the zero pole, and so\footnote{The zero pole contribution in this case gives two times the polynomial terms. This factor of two is discussed in detail in \cite{Hattab:2024chf}.}
\be 
I(\tilde{u})^{\mathrm{ResCo}} =  \frac{2\pi}{i\tilde{u}^2}\frac{1}{(2\sin(\pi \tilde{u}))^{2}}  \;, \;\; \psi(\mu)=e^{\mu} \;.
\ee 


The case of a compact Calabi-Yau (with strict $SU(3)$ holonomy) is the most involved. The resummation is captured by mirror symmetry, which can be a very complicated map. For example, for the compact one-parameter Calabi-Yau studied in \cite{Joshi:2019nzi,Palti:2021ubp,Bastian:2023shf,Hattab:2023moj}, we have \cite{Hattab:2024thi}:
\begin{eqnarray}
& & I(\tilde{u})^{\mathrm{CY}} = \;\frac{9}{2\pi i}3^{6\tilde{u}}\Gamma(-\tilde{u})^4 \left[\frac{ \Gamma\left(\frac13 + \tilde{u}\right)^2 }{\Gamma\left(\frac13 - \tilde{u}\right)^2}   \nonumber \right. \\ \nonumber
& & +\frac{ \Gamma\left(\frac23 + \tilde{u}\right)^2 }{ \Gamma\left(\frac23 - \tilde{u}\right)^2}
 - \frac{ (-1)^{-\tilde{u}} (3i+\sqrt{3})\pi \Gamma\left(\frac23 + \tilde{u}\right) }{\Gamma\left(\frac13 - \tilde{u}\right) \Gamma(\frac23 - \tilde{u})^2} \nonumber \\
& & \left. +\frac{(-1)^{-\tilde{u}} (3i - \sqrt{3})\pi \Gamma\left(\frac13 + \tilde{u}\right) }{\Gamma\left(\frac13 - \tilde{u}\right)^2 \Gamma\left(\frac23 - \tilde{u}\right)}\right] \;,
\label{granp5}
\end{eqnarray}
with
\be 
\label{apprps}
\psi(\mu) = e^{\mu} + ...  \;.
\ee 
In this case, we have that the zero pole in $\tilde{u}$ yields both polynomial and exponential terms in $\mu$, coming from the sub-leading terms in (\ref{apprps}).

\section{Back to weak coupling}
\label{sec:backweak}

After completing the integrating out calculation in the strongly-coupled regime, so at leading order in $\hbar$, we would like to connect back to the weakly-coupled regime. The coordinate transformation back from $\mu$ to $t$, given by (\ref{Ttomu}), does not maintain an order-by-order match in the expansions (\ref{Fexp}) and (\ref{Fmuexpan}). Nonetheless, some of the terms in $F_0(t)$ and $F_1(t)$ can be deduced from $J_0(\mu)$. This was explained already in \cite{Hattab:2024thi}, and we simply recall it here for completeness. 

The zero pole of $J_0(\mu)$ yields, in general, polynomial terms of the form
\be
\label{j0polyst}
J_0(\mu) \supset -\frac{1}{6}\kappa \mu^3+\frac{\pi^2}{6}c_2\;\mu +\chi(Y)\zeta(3)\;.
\ee
At strong coupling, $\kappa$, $c_2$ and $\chi(Y)$ are just coefficients, but they have geometric interpretations at weak coupling: $\kappa$ is the triple self-intersection of the Kahler form, $c_2$ is the intersection of the second Chern class $c_2(Y)$ with the Kahler form, and $\chi(Y)$ the Euler characteristic of the Calabi-Yau $Y$. 

 Using the coordinate map (\ref{Ttomu}) on (\ref{j0polyst}) we obtain
\begin{eqnarray}
     \frac{1}{2\pi \hbar}\left(-\frac{1}{6}\kappa \mu^3\right)&=& -\frac{1}{\lambda^2}\frac{1}{6}\kappa \;t^3 \;, \label{rel1} \\
    \frac{1}{2\pi \hbar}\left(\frac{\pi^2}{6}c_2\;\mu\right) &=&  \frac{c_2}{24} t \;\;.  \label{rel2}
\end{eqnarray}
We therefore obtain the cubic piece in $F_0(t)$ and the linear piece in $F_1(t)$. These are the leading terms at their given order in $\lambda$ and $t$. We therefore have shown that the leading classical terms in the free energy, or the supergravity effective action, arise from integrating out M2 states. 

Note that the relations (\ref{rel1}) and (\ref{rel2}), together with the fact that the linear term in $F_0$ is $-\left(2\pi\right)^2$ times the linear term in $F_1$, show the polynomial part of the equality (\ref{cyper}).

The constant piece $\chi(Y)\zeta(3)/2\pi \hbar$ is part of the non-perturbative completion of the D0 integrating out formula, given by setting $\beta=0$ in (\ref{j0intrep}), so
\begin{eqnarray}
   -\frac{\chi(Y)}{2}\oint_{C}\frac{du}{u}\frac{1}{1-e^{-2\pi i u}}\frac{1}{(2\sin(u\lambda/2))^2} \;,
\end{eqnarray}
which yields a contribution at leading order in $\hbar$ of
\be 
\chi(Y)\zeta(3)  \in J_0(\mu)\;,
\ee 
and at leading order in $\lambda$ of 
\be 
-\frac12 \chi(Y)\zeta(3)  \in F_0(t)\;.
\ee

\section{Summary}
\label{sec:sum}

In this note we studied the strong-coupling behaviour of the free energy as given in (\ref{full}). We showed that it exhibits a strong-weak coupling duality relation (\ref{cyper}). This was conjectured to hold, for general Calabi-Yau manifolds, in \cite{Hattab:2024thi}. It also previously appeared in non-compact examples \cite{Drukker:2010nc,Marino:2011eh,Grassi:2014zfa}. 

The relation (\ref{cyper}) yields a prescription for completing the integrating out calculation in the ultraviolet. In particular, including the ultraviolet contribution this way shows that the leading classical terms in the free energy, and the supergravity effective action, emerge from integrating out M2 states at one-loop, as proposed by Emergence. This was already proposed and developed in \cite{Hattab:2023moj,Hattab:2024thi,Hattab:2024ewk,Hattab:2024chf}, but we presented some additional details here. 

It is important to note that the relation between integrating out states and the leading classical terms holds true regardless of whether (\ref{full}) captures the full non-perturbative topological string free energy. It is the full result of the integrating out calculation \cite{Hattab:2024ewk}, and that is all that is utilised in this aspect of the analysis. 

It is interesting to note that instead of expanding in small $\hbar$, we could consider the `self-dual' point
\be 
\lambda = \hbar = 2\pi \;.
\ee 
 We see from (\ref{full}) that this is the point where the two types of poles coincide perfectly, and also where all higher genus $r \geq 2$ contributions vanish. The free energy then simplifies considerably. It would be interesting to understand the physics of this self-dual point.\footnote{
In the case of the blown-up conifold, this is the `maximally symmetric' point where the dual ABJM theory \cite{Aharony:2008ug} enhances to ${\cal N}=8$ supersymmetry \cite{Codesido:2014oua}.}

\vspace{0.3cm}
{\bf Acknowledgements}
\noindent
The work of JH and EP is supported by the Israel Science Foundation (grant No. 741/20) and by the German Research Foundation through a German-Israeli Project Cooperation (DIP) grant ``Holography and the Swampland". The work of EP is supported by the Israel planning and budgeting committee grant for supporting theoretical high energy physics.

\appendix

\section{The particle picture}
\label{sec:part}

One can also derive the non-perturbative formula (\ref{full}) using a particle picture approach. The expression (\ref{full}) is the result of a field theory integrating out calculation after KK reduction of the five-dimensional M-theory background on the M-theory circle. As explained in \cite{Dedushenko:2014nya}, the right supersymmetric background to perform the integrating-out calculation is a constant anti-self-dual graviphoton field strength. To identify the compactified time coordinate with the M-theory circle one needs to Wick rotate to Euclidean space, but for the metric to stay real the graviphoton field strength needs to become pure imaginary. This is why we have a $\sin(\lambda)$ in the denominator of (\ref{full}) instead of the usual $\sinh(\lambda)$. The energy levels of an M2 particle excitation in a constant anti-self dual field strength in $\mathbb{R}^4$ are two copies of the same system in $\mathbb{R}^2$ which has usual Landau levels $E_n = -i\lambda (n+\frac{1}{2})$, where $-i\lambda$ takes into account the fact that the Field strength needs to be purely imaginary (see \cite{Dedushenko:2014nya} for the detailed calculation in the Type IIA side). This means that an M2 particle excitation of mass $t$ winding the M-theory circle $k$ times will contribute
\be
-\frac{e^{-kt}}{k}\text{Tr}\left(e^{-k \hat{H}}\right) = -\frac{e^{-kt}}{k}\left(\sum_{n\geq 0}e^{ik\lambda(n+\frac{1}{2})}\right)^2\;.
\ee
It is clear here that summing over the energy eigenstates is not possible since the sum is not convergent. The right way to do this is to sum over the windings first. In some sense, this amounts to treating the different M2 excitations as part of the same objects and not simply as independent particles. The sum over k yields
\begin{eqnarray}
    &&-\sum_{n ,m\geq  0}\sum_{k\geq 1}\frac{e^{-kt}}{k}e^{i k\lambda(n+m+1)} \nonumber \\
    &=& \sum_{m,n\geq 0}\log(1-e^{-t+i\lambda(m+n+1)}) \nonumber \\
    &=& \log\left[\prod_{m,n\geq 0}\left(1-e^{-t+i\lambda(m+n+1)}\right)\right] \nonumber \\
    &=& \oint_C\frac{\text{d}u}{u}\frac{1}{1-e^{-2\pi i u}}\frac{e^{-tu}}{\big(2\sin(u\lambda/2)\big)^{2}} \;.
    \label{intuparpic}
\end{eqnarray}
In the last line we used that, for $\lambda\in \mathbb{R}^+$ and $\text{Im}(t)\in(0,2\pi)$, the infinite product is a representation of a function closely related to the triple sine function \cite{narukawa2004modularpropertiesintegralrepresentations}, that can be written through a contour integral representation \cite{Bridgeland:2017vbr,narukawa2004modularpropertiesintegralrepresentations}.\footnote{This function was already identified in \cite{Krefl:2015vna,Bridgeland:2017vbr,Alim:2021ukq} as the non-perturbative completion for the resolved conifold through multiple different methods.}
The expression (\ref{intuparpic}) holds for a single particle, and so (\ref{full}) follows simply from summing over the different particles.


\bibliography{Higuchi}

\end{document}